# Unsteady Hydromagnetic Flow Of Viscoelastic Fluid Down An Open Inclined Channel

S.Sreekanth1,, R.Saravana2, S.Venkataramana3, R.Hemadri Reddy4
1Department of Mathematics, Sreenivasa Institute of Technology and Management Studies, Chittoor 517127, A.P. India.
2 & 3Department of Mathematics, Sri Venkateswara University, Tirupati, A.P. India.
4School of Advanced Sciences, VIT University, Vellore – 632 014, T.N. India.

**Abstract-In this paper, we study the unsteady hydromagnetic flow of a Walter's fluid (Model B') down an open inclined channel of width 2a and depth d under gravity, the walls of the channel being normal to the surface of the bottom under the influence of a uniform transverse magnetic field. A uniform tangential stress is applied at the free surface in the direction of flow. We have evaluated the velocity distribution by using Laplace transform and finite Fourier Sine transform technique. The velocity distribution has been obtained taking different form of time dependent pressure gradient g(t), viz., i) constant ii) exponential decreasing function of time and iii) Cosine function of time. The effects of magnetic parameter M, Reynolds number R and the viscoelastic parameter K are discussed on the velocity distribution in three different cases.**

## 1. INTRODUCTION

A flowing liquid is said to have a free surface when the upper part of the bounding surface of the liquid is in contact with the overlying atmosphere, rather than with a solid, as would be the case of the flow were in a pipe completely full of the liquid. A flow with a free surface proceeding in a natural or artificial channel or conduit is called an open-channels flow. Open-channels may be divided into two types namely (i) Natural channels and (ii) Artificial channels. Natural channels range inform from the bounder-strewn bed of a mountain torrent to the relatively uniform channel of a large river. Artificial channels are man-made and are constructed in many forms. A pipe in which water flows with a free surface and flume of rectangular cross section constructed of sheet iron are two kinds of artificial channels. Other types are canals excavated in earth or blasted in rock, which either are left unlined or lined with smooth concrete or another suitable material. Unlined or lined tunnels bored through rock may also contain free surface flows.

The flow of a liquid in an open inclined channel with a free surface has a wide application in the designs of drainage, irrigation canals, flood discharge channels and coating to paper rolls etc. Hence the flow of a liquid in an open inclined channel with a free surface under gravity has long been studied experimentally and several interesting empirical results have been reported by many investigators [3, 6, 7, 10, 11, 14]. The steady laminar flow of a viscous fluid flowing down an open inclined channel has been discussed by Satyaprakash [13], Gupta *et al* [4] have studied the flow of a viscous fluid through a porous medium down an open inclined channel. Venkataramana and Bathaiah [18] have studied the flow of a hydromagnetic viscous fluid down an open inclined channel with naturally permeable bed under the influence of a uniform transverse magnetic field. Unsteady laminar flow of an incompressible viscous fluid between porous, parallel flat plates has been investigated by Singh [12], taking (i) both plates are at rest and (ii) Generalized plane Coutte flow. The free surface was exposed to atmospheric pressure and bottom was taken as impermeable. Bakhmeteff [1], Henderson [5] and Chow [2] have discussed many types of open channel flows. Recently, many authors [8, 9 and 14] have studied the flow of Walter's B' fluid.

The subject of Rheology is of great technological importance in many branches of industry. The problem arises of designing apparatus to transport or to process substances which cannot be governed by the classical stress-strain velocity relations. Example of such substances and process are many, the extrusion of plastics, in the manufacture of rayon, nylon or other textile fibres, viscoelastic effects transported or forced through spinnerts and in the manufacture of lubricating grease and rubber.

Non-Newtonian fluids have wide importance in the present day technology and industries; the Walter's fluid is one of such fluid. The model of Walter's B fluid is chosen for our study as it involves non-Newtonian parameter. The Cauchy stress tensor *T* in such a fluid is related to the motion in the following manner

$$T = -PI + 2\eta_0 e - 2K_0 \frac{\delta e}{\delta t} \qquad (1.1)$$

In this equation *P* is the pressure, *I* is the Identity tensor and the rate of strain tensor e is defined by

$$2e = \nabla v + (\nabla v)^T \qquad (1.2)$$

where *v* is the velocity vector, $\nabla$ is the gradient operator and $\frac{\delta}{\delta t}$ denotes the convicted differentiation of a tensor quantity in relation to the material in motion. The convicted differentiation of the rate of strain tensor is given by

$$\frac{\delta e}{\delta t} = \frac{\partial e}{\partial t} + v \cdot \nabla e - e \cdot \nabla v - (\nabla v)^T \cdot e \qquad (1.3)$$

Here $\eta_0$ and $K_0$ are, respectively, the limiting viscosity at small rate of shear and the short memory coefficient which are defined through

$$\eta_0 = \int_0^\infty N(\tau)\, d\tau \qquad (1.4)$$

and

$$K_0 = \int_0^\infty \tau N(\tau)\, d\tau \qquad (1.5)$$

$N(\tau)$ being the relaxation spectrum as introduced by Walter's [19, 20]. This idealized model is a valid approximation of Walter's fluid (model B') taking very short memory into account, so that terms involving

$$\int_0^\infty \tau^n N(\tau)\, d\tau, \qquad n \geq 2 \qquad (1.6)$$

have been neglected.

In addition to equation (1.1), the equation of motion and continuity are

$$\nabla \cdot e = 0 \qquad (1.7)$$

$$\rho (v \cdot \nabla v) = \nabla \cdot T \qquad (1.8)$$

In this paper, we study the unsteady hydromagnetic flow of a Walter's fluid (model B') down an open inclined channel under gravity of width 2a and depth d, the walls of the channel being normal to the surface of the bottom, under the influence of uniform transverse magnetic field. A uniform tangential stress is applied at the free surface in the direction of flow. We have evaluated the velocity distribution by using Laplace Transform and Finite Fourier Sine Transform techniques. Here it is assumed that (i) the fluid flows in the steady state for $t \leq 0$, (ii) Unsteady state occurs at $t > 0$ and (iii) the unsteady motion is influenced by time dependent pressure gradient. The velocity distribution has been obtained in some particular cases i.e. when (i) $g(t) = c^*$ (ii) $g(t) = c^* e^{-bt}$ and (iii) $g(t) = c^* \cos bt$, where b and $c^*$ are constants. The effects of magnetic parameter M, Reynolds number R and viscoelastic parameter K are investigated on the velocity distribution in three different cases.

## 2. FORMULATION AND SOLUTION OF THE PROBLEM

We consider the unsteady Hydromagnetic flow of a Walter's fluid (model B$'$) down an open inclined channel of width 2a and depth d under gravity, the walls of the channel being normal to the surface of the bottom under the influence of uniform transverse magnetic field. A uniform tangential stress S is applied at the free surface. The bottom of the channel is taken at angle $\beta\left(0<\beta\leq\pi/2\right)$ with the horizontal. The x-axis is taken along central line in the direction of the flow at the free surface, y-axis along the depth of the channel and z-axis along width of the channel. A uniform magnetic field of intensity $H_o$ is introduced in y-direction. Therefore the velocity and the magnetic field are given by $\overline{q}=\left(u,0,0\right)$ and $\overline{H}=\left(0,H_0,0\right)$. The fluid being slightly conducting, the magnetic Reynolds number is much less than unity, so that the induced magnetic field can be neglected in comparison with the applied magnetic field (Sparrow and Cess [15]). In the absence of any input electric field the equations of continuity and motion of the unsteady hydromagnetic Walter's fluid (model B$'$) flowing down an open inclined channel at t>0 are given by

$$\frac{\partial u}{\partial x}=0 \tag{2.1}$$

$$\rho\frac{\partial u}{\partial t}=-\frac{\partial p}{\partial X}+\rho g\sin\beta+\mu\left(\frac{\partial^2 u}{\partial y^2}+\frac{\partial^2 u}{\partial z^2}\right)-K_0\left(\frac{\partial^3 u}{\partial t\partial y^2}+\frac{\partial^3 u}{\partial t\partial z^2}\right)-\sigma\mu_e^{\,2}H_0^{\,2}u \tag{2.2}$$

$$0=-\frac{\partial p}{\partial y}+\rho g\cos\beta \tag{2.3}$$

$$0=-\frac{\partial p}{\partial z} \tag{2.4}$$

Where $\rho$ = density of the fluid

g = acceleration due to gravity

p = pressure

$\mu$ = coefficient of viscosity

$\sigma$ = electrical conductivity of the fluid

$\mu_e$ = magnetic permeability

$K_0$ = viscoelastic parameter

The boundary conditions are

$$\left.\begin{array}{l} t \le 0;\ u = u_0 \\ t > 0;\ z = \pm a, u = o \\ y = 0,\ \mu \dfrac{\partial u}{\partial y} = S \\ y = d\ \ u = 0 \end{array}\right\} \tag{2.5}$$

where $u_0$ is the initial velocity.

we introduce the following non-dimensional quantities

$$\left.\begin{array}{l} u^* = u / U, \quad x^* = x / d, \quad y^* = y / d, \quad z^* = z / d, \quad t^* = t\upsilon / d^2 \\ p^* = p / \rho U^2, \quad K^* = K_0 / \rho d^2, S^* = S / \rho U^2 \end{array}\right\} \tag{2.6}$$

where the depth of the channel d is the characteristic length and the mean flow velocity U is the characteristic velocity.

In view of the equation (2.6), the equations (2.1) to (2.3) reduce to (dropping the superscript *)

$$\frac{\partial u}{\partial t} = -R\frac{\partial P}{\partial X} + \frac{R}{F}\sin\beta + \frac{\partial^2 u}{\partial y^2} + \frac{\partial^2 u}{\partial z^2} - K\left(\frac{\partial^3 u}{\partial t \partial y^2} + \frac{\partial^3 u}{\partial t \partial z^2}\right) - Mu \tag{2.7}$$

$$\frac{\partial u}{\partial x} = 0 \tag{2.8}$$

Where $M = \sigma \mu_e^{\ 2} H_o^{\ 2} d^2 / \rho\upsilon$ Magnetic parameter

$R = Ud / \upsilon$ Reynolds number

$F = U^2 / gd$ Froude number

$K = K_0 / \rho d^2$ viscoelastic parameter

The non-dimensional boundary conditions are

$$\left.\begin{array}{l} t \le 0;\ u = u_0 \\ t > 0;\ z = 1(= \pm a / d), u = o \\ y = 0\ \partial u / \partial y = SR \\ y = 1\ u = 0 \end{array}\right\} \tag{2.9}$$

### 3. METHOD OF SOLUTION

Assuming

$$-R\frac{\partial p}{\partial X}+\frac{R}{F}\sin\beta \qquad = g(t) \quad at \quad t>0$$
$$= P \qquad at \quad t\le 0 \tag{3.1}$$

substituting z = (2lf/π)-1 in equation (2.7) reduces to

$$\frac{\partial u}{\partial t}=g(t)+\frac{\partial^2 u}{\partial y^2}+\frac{\pi^2}{4l^2}\frac{\partial^2 u}{\partial f^2}-K\left(\frac{\partial^2 u}{\partial t\partial y^2}+\frac{\pi^2}{4l^2}\frac{\partial^3 u}{\partial t\partial f^2}\right)-Mu \tag{3.2}$$

and the boundary conditions are reduced to

$$\left.\begin{array}{l} t\le 0;\ u=u_0 \\ t>0;\ \ f=0,\pi;\ \ u=o \\ y=0,\ \dfrac{\partial u}{\partial y}=SR \\ y=1,\ u=0 \end{array}\right\} \tag{3.3}$$

Now, since $u_0$ is the initial velocity i.e. at $t\le 0$, therefore taking $g(t)=P$ in equation (3.2)

$$u_0=\frac{2}{\pi}\sum_{n=1}^{\infty}\left(\frac{1-\cos n\pi}{n}\right)\left[\frac{P}{C^2}\left(1-\frac{\cosh Cy}{\cosh C}\right)-\frac{SR}{C}\frac{\sinh C(1-y)}{\cosh C}\right]\sin nf \tag{3.4}$$

Where

$$C^2=Q^2+M, \qquad\qquad Q=\frac{n\pi}{2l}$$

Now to solve equation (3.2) we take Laplace transform of equation (3.2) with respect to t (Sneddon [16])

$$\bar{u}(y,f,s)=\int_0^\infty u(y,f,s)\,e^{-st}\,dt, \qquad\qquad s>0 \tag{3.5}$$

we get

$$\frac{\partial^2\bar{u}}{\partial y^2}+\frac{\pi^2\partial^2\bar{u}}{4l^2\partial f^2}-\frac{(M+s)}{1-Ks}\bar{u}=\frac{1}{1-Ks}\left(\frac{pKM}{C^2}\frac{\cosh Cy}{\cosh C}+\frac{PKQ^2}{C^2}+\frac{SRMK}{C}\frac{\sinh C(1-y)}{\cosh C}-\bar{g}(s)-u_0\right)$$

(3.6)

Where $\bar{g}(s) = \int_0^\infty g(t)\, e^{-st}\, dt$

On taking the finite Fourier sine transform the equation (3.6) with respect to f (Sneddon [16])

$$\bar{u}^*(y, N, s) = \int_0^\pi \bar{u}(y, f, s) \sin Nf df \qquad (3.7)$$

We get

$$\frac{\partial^2 \bar{u}^*}{\partial y^2} - H^2 \bar{u}^* = \left(\frac{1-\cos n\pi}{N(1-Ks)}\right)\left(\frac{PKQ^2}{C^2} - \bar{g}(s) - \frac{P}{C^2} + \frac{P(1+MK)}{C^2}\frac{\cosh Cy}{\cosh C} + \frac{SR(1+MK)}{C}\frac{\sinh Cy}{\cosh C}\right)$$

(3.8)

Where $H^2 = Q^2 + \dfrac{s+M}{1-Ks}$

Now, the boundary conditions are reduced to

$$\left.\begin{aligned} y=0, \quad & \frac{\partial \bar{u}^*}{\partial y} = \frac{SR(1-\cos N\pi)}{SN} \\ y=1, \quad & \bar{u}^* = 0 \end{aligned}\right\} \qquad (3.9)$$

Integrating equation (3.8) under the boundary conditions (3.9), we get

$$\bar{u}^* = \left(\frac{1-\cos N\pi}{N}\right)\left(\begin{aligned} & \frac{PKQ^2}{C^2H^2(1-Ks)}\left(\frac{\cosh Hy}{\cosh H}-1\right) + \frac{g(\bar{s})}{H^2(1-Ks)}\left(\frac{\cosh Hy}{\cosh H}-1\right) - \frac{P}{H^2C^2(1-Ks)}\left(\frac{\cosh Hy}{\cosh H}-1\right) \\ & + \frac{P}{SC^2}\left(\frac{CoshHy}{\cosh H} - \frac{\cosh Cy}{\cosh C} - \frac{SR}{SC}\frac{\sinh C(1-y)}{\cosh C}\right) \end{aligned}\right)$$

(3.10)

Now, inverting the finite Fourier sine transform as given by (Sneddon [16])

$$\bar{u}(y, f, s) = \frac{2}{\pi}\sum_{N=1}^{\infty} \bar{u}^*(y, N, s) \sin Nf$$

In equation (3.10) we get

$$\bar{u}(y,f,s)=\frac{2}{p}\sum_{N=1}^{\infty}\left(\frac{1-\cos Np}{N}\right)\left[\frac{PKQ^2}{C^2H^2(1-Ks)}\left(\frac{\cosh Hy}{\cosh H}-1\right)-\frac{g(s)}{H^2(1-Ks)}\left(\frac{\cosh Hy}{\cosh H}-1\right)\right.$$
$$\left.-\frac{P}{H^2C^2(1-Ks)}\left(\frac{\cos Hy}{\cosh H}-1\right)+\frac{P}{SC^2}\left(\frac{\cosh Hy}{\cosh H}-\frac{\cosh Cy}{\cosh C}\right)-\frac{SR}{SC}\frac{\sinh C(1-y)}{\cosh C}\right]\sin Nf \tag{3.11}$$

On inverting Laplace transform as defined by (Sneddon[16])

$$u(y,f,t)=\frac{1}{2pi}\int_{r-i\infty}^{r+i\infty}\bar{u}(y,f,s)e^{st}\,dt$$

In equation (3.11), we obtain

$$u=\frac{2}{\pi}\sum_{N=1}^{\infty}\left(\frac{1-\cos N\pi}{N}\right)\left[\sum_{r=0}^{\infty}\frac{4P(-1)^r e^{-A_r t}\cos a_r y}{\pi(2r+1)\left(a_r^{\,2}+c^2\right)}+\int_0^t h(u)\,g(t-u)\,du-\frac{SR}{C}\frac{\sinh C(1-y)}{\cosh C}\right]\sin Nf \tag{3.12}$$

Where $$h(u)=\sum_{r=0}^{\infty}\frac{4(-1)^r\cos aye^{-A_r u}}{\pi(2r+1)\left\{1-K\left(a_r^{\,2}+Q^2\right)\right\}},\quad a_r=\frac{\pi}{2}(2r+1)$$

and $$A_r=\frac{\left(a_r^{\,2}+c^2\right)}{1-k\left(a_r^{\,2}+Q^2\right)}$$

**Particular cases**

**Case 1.** When $g(t)=C^*$

Using in equation (3.12), we get

$$u=\frac{2}{\pi}\sum_{N=1}^{\infty}\left(\frac{1-\cos N\pi}{N}\right)\left[\sum_{r=0}^{\infty}\frac{4(-1)^r e^{-A_r t}\cos a_r y}{\pi(2r+1)\left(a_r^{\,2}+c^2\right)}(P-C^*)+\frac{C^*}{C^2}\left(1-\frac{\cosh Cy}{\cosh C}\right)\right.$$
$$\left.-\frac{SR}{C}\frac{\sinh C(1-y)}{\cosh C}\right]\sin Nf \tag{3.13}$$

If we take the limit $M\to 0, K\to 0,\ g(t)=P=C^*$ in equation (3.13) then we get the velocity distribution in the case of nonmagnetic and Newtonian fluid. In this case the velocity distribution is

$$u=-\frac{2}{\pi}\sum_{N=1}^{\infty}\left(\frac{1-\cos N\pi}{N}\right)\left[\frac{C}{Q^2}\left(1-\frac{\cosh Qy}{\cosh Q}\right)-\frac{SR}{Q}\frac{\sinh Q(1-y)}{\cosh Q}\right]\sin Nf \tag{3.14}$$

This is in agreement with Sathyaprakash [13]

**Case II**

When $g(t)=C^{*}e^{-bt},\ b>0$, using in equation (3.12), we get

$$u=\frac{2}{\pi}\sum_{N=1}^{\infty}\left(\frac{1-\cos N\pi}{N}\right)[\sum_{r=0}^{\infty}\frac{4P(-1)^r\cos a_r y e^{-A_r t}}{\pi(2r+1)\left(a_r^{\ 2}+C^2\right)}$$

$$+\sum_{r=0}^{\infty}\frac{4C^{*}(-1)^r\cos a_r y e^{-bt}-e^{-A_r t}}{\pi(2r+1)\left\{1-K\left(a_r^{\ 2}+Q^2\right)\right\}(A_r-b)}-\frac{SR}{C}\frac{\sinh C(1-y)}{\cosh C}]\sin Nf \tag{3.15}$$

**Case III**

When $g(t)=C^{*}\cos bt$, using in equation (3.12), we get

$$u=\frac{2}{\pi}\sum_{N=1}^{\infty}\left(\frac{1-\cos N\pi}{N}\right)[\sum_{r=0}^{\infty}\frac{4P(-1)^r\cos a_r y e^{-A_r t}}{\pi(2r+1)\left(a_r^{\ 2}+C^2\right)}$$

$$+\sum_{r=0}^{\infty}\frac{4C^{*}(-1)^r\cos a_r y\left(A_r\cos bt+b\sin bt-A_r e^{-A_r t}\right)}{\pi(2r+1)\left\{1-K\left(a_r^{\ 2}+Q^2\right)\right\}\left(A_r^{\ 2}+b^2\right)}-\frac{SR}{C}\frac{\sinh C(1-y)}{\cosh C}]\sin Nf$$

(3.16)

## 4. CONCLUSIONS

Three distinct time dependent pressure gradient namely (i) $g(t)=C^{*}$ (ii) $g(t)=C^{*}e^{-bt}$ and (iii) $g(t)=C^{*}\cos bt$ have been chosen to discuss the velocity profiles. Figures (1) to (3), (4) to (6), (7) to (9) are drawn to investigate the effects of magnetic parameter M, time t and viscoelastic parameter K on the velocity distribution u in three different cases (i), (ii), (iii) respectively. In all the three cases we noticed that the velocity distribution increases with the increase in M or t where as it decreases with increase in K. Further we observe that in all the three cases the velocity profiles in nonmagnetic case are greater than with the magnetic case. Table (1) is drawn to bring out the effects of Reynolds number R on velocity distribution in three different cases. We noticed that in all the three cases the velocity distribution decreases with the increase in R. Further it is observed that in all three cases the velocity profiles in nonmagnetic case are less than with the magnetic case.

**GRAPHS**

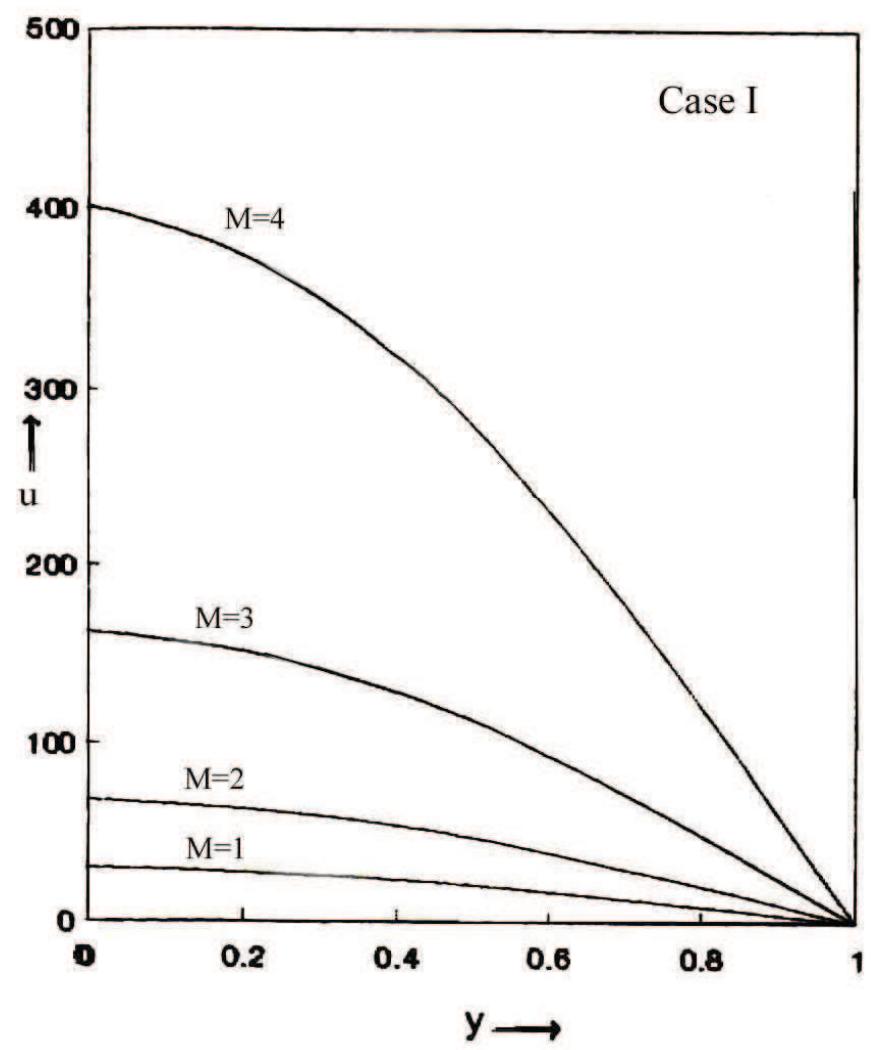

Fig.1 u against y for different M

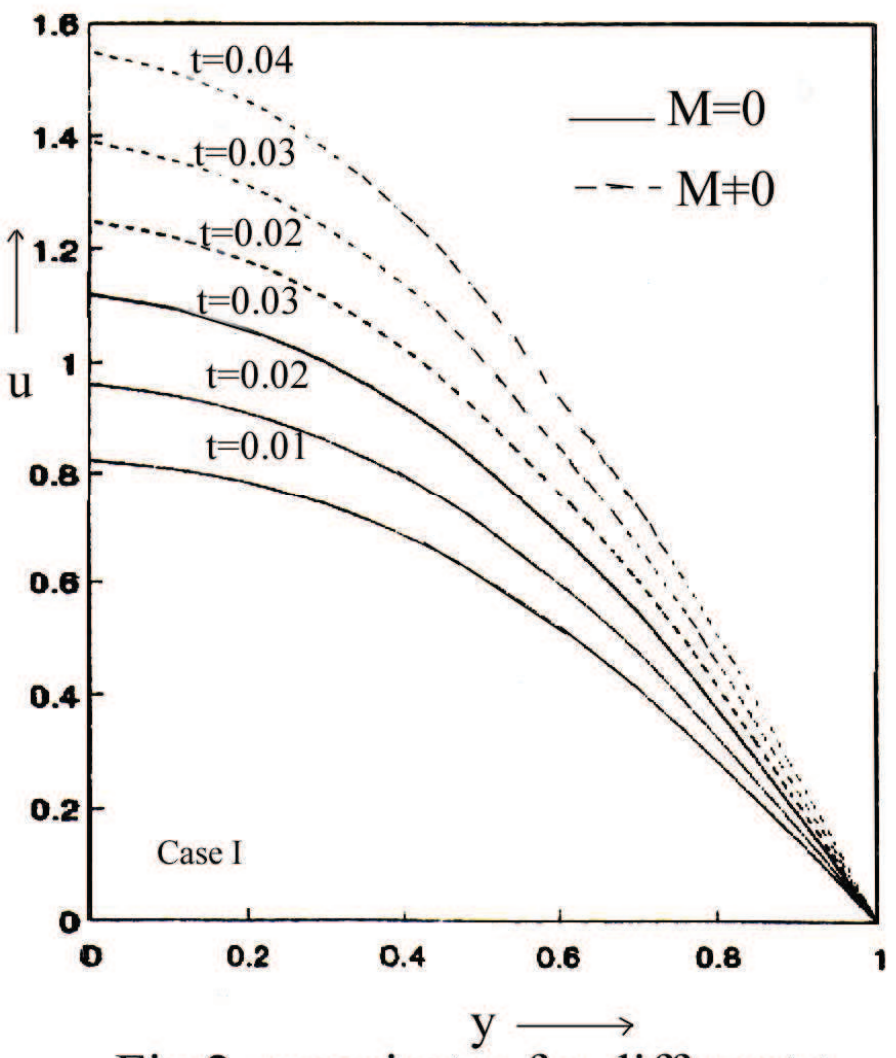

Fig.2 u against y for different t

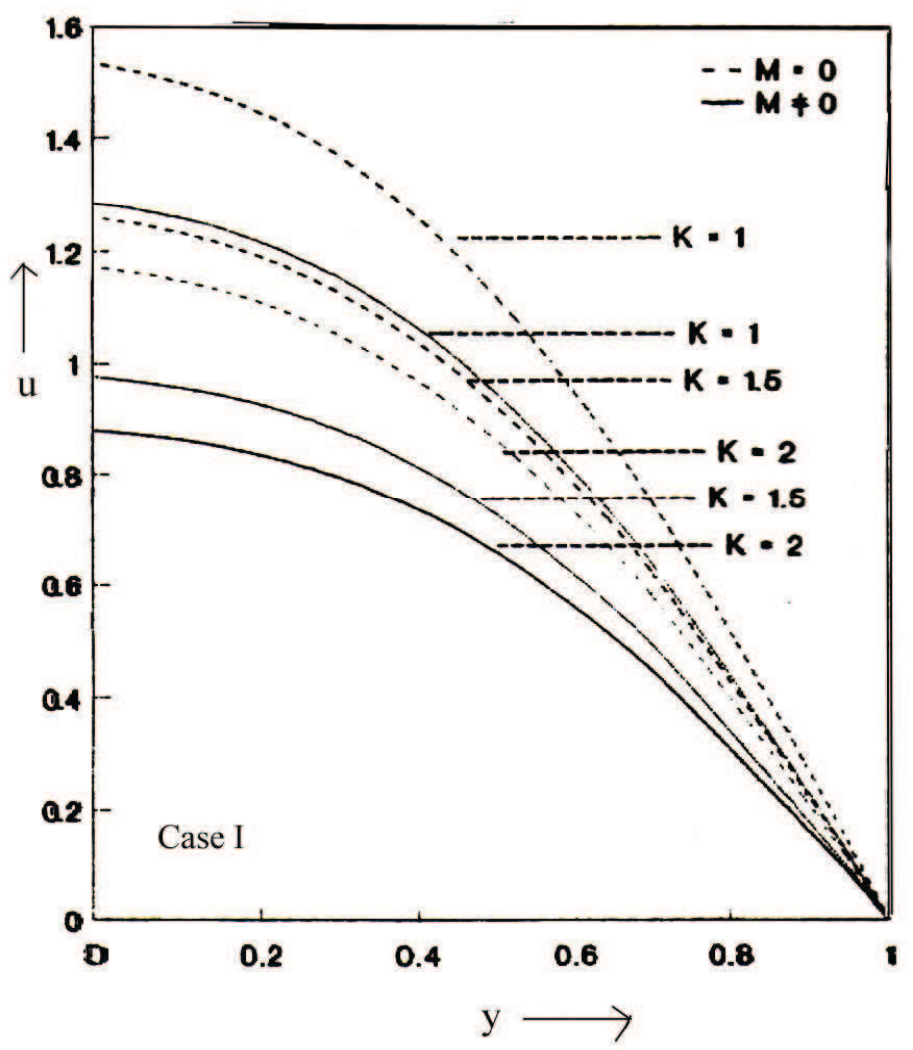

Fig.3 u against y for different K

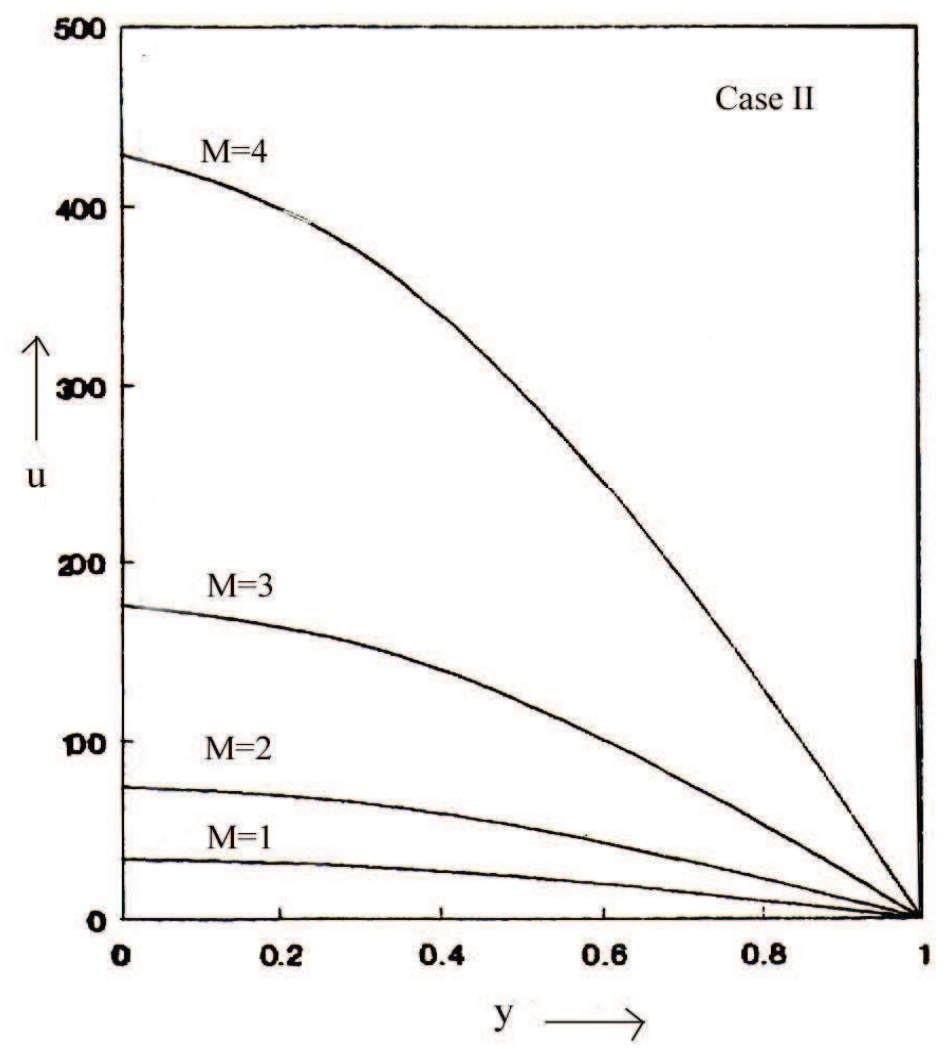

Fig.4 u against y for different M

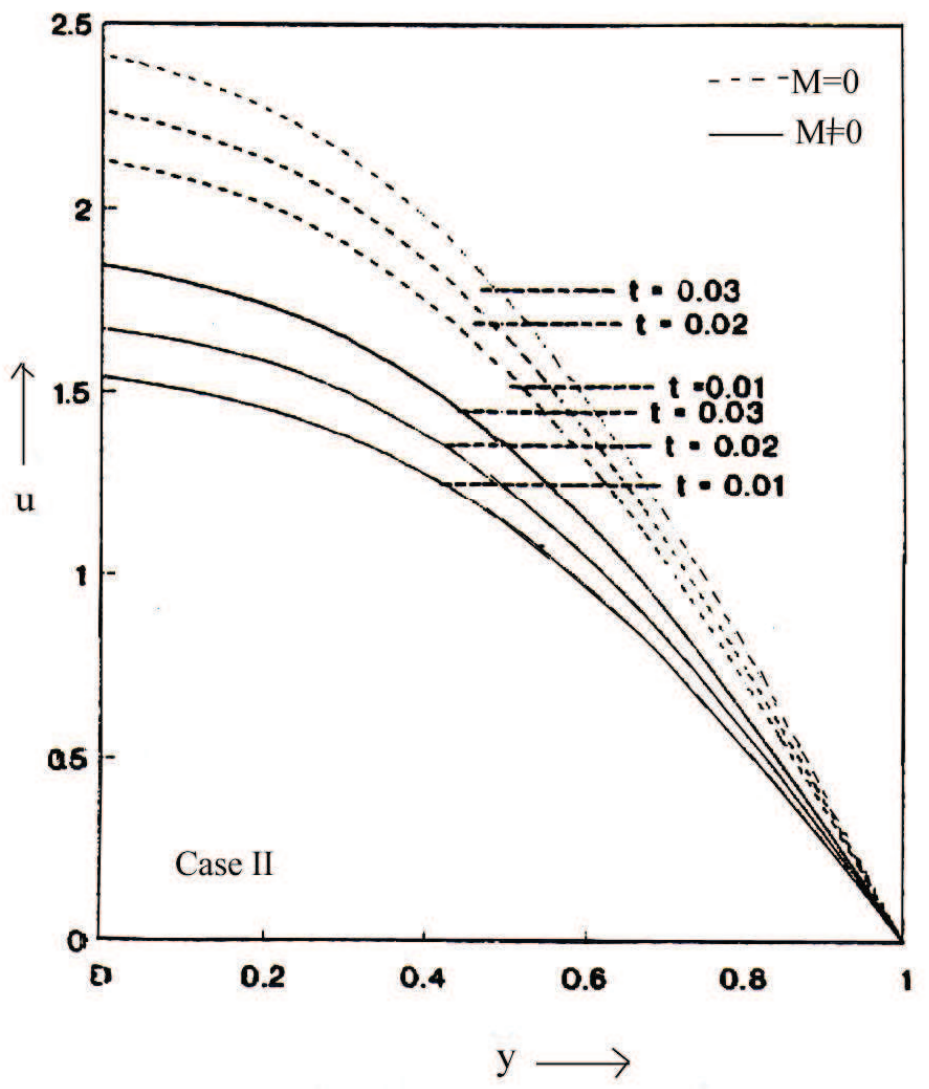

Fig.5 u against y for different t

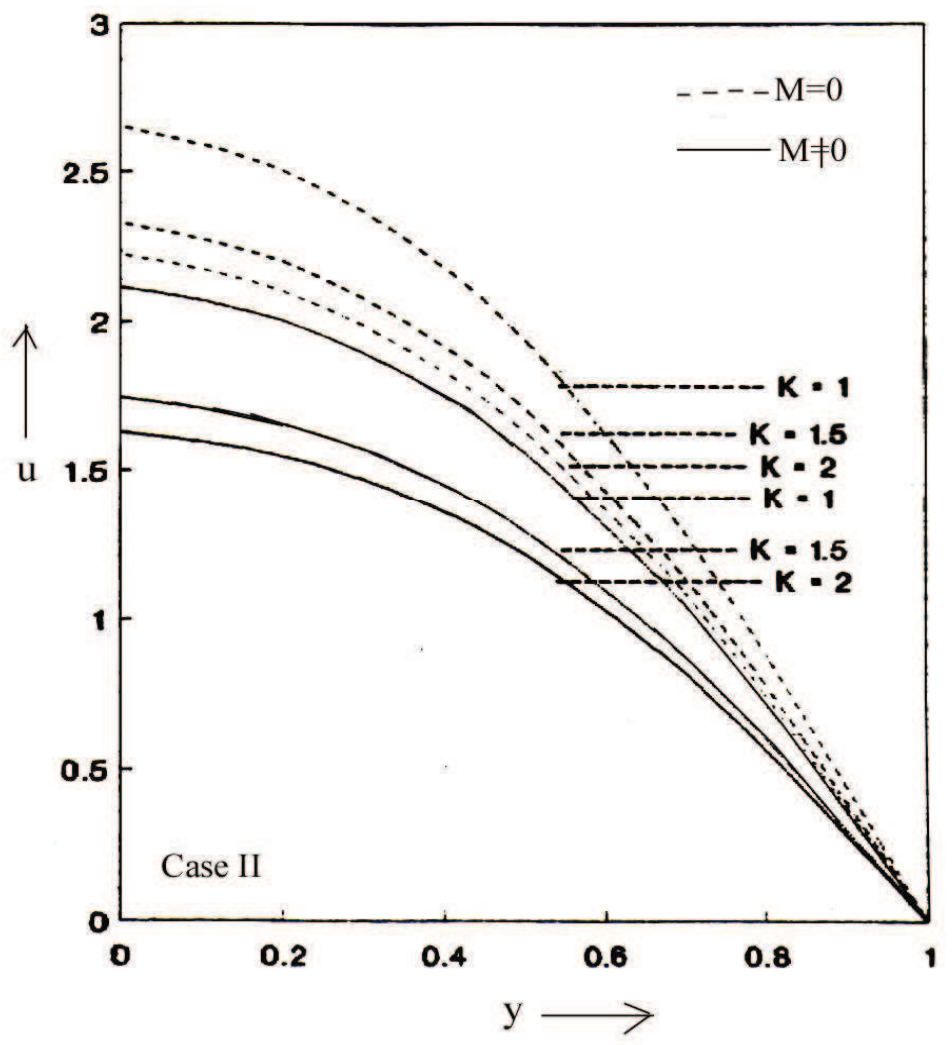

Fig.6 u against y for different K

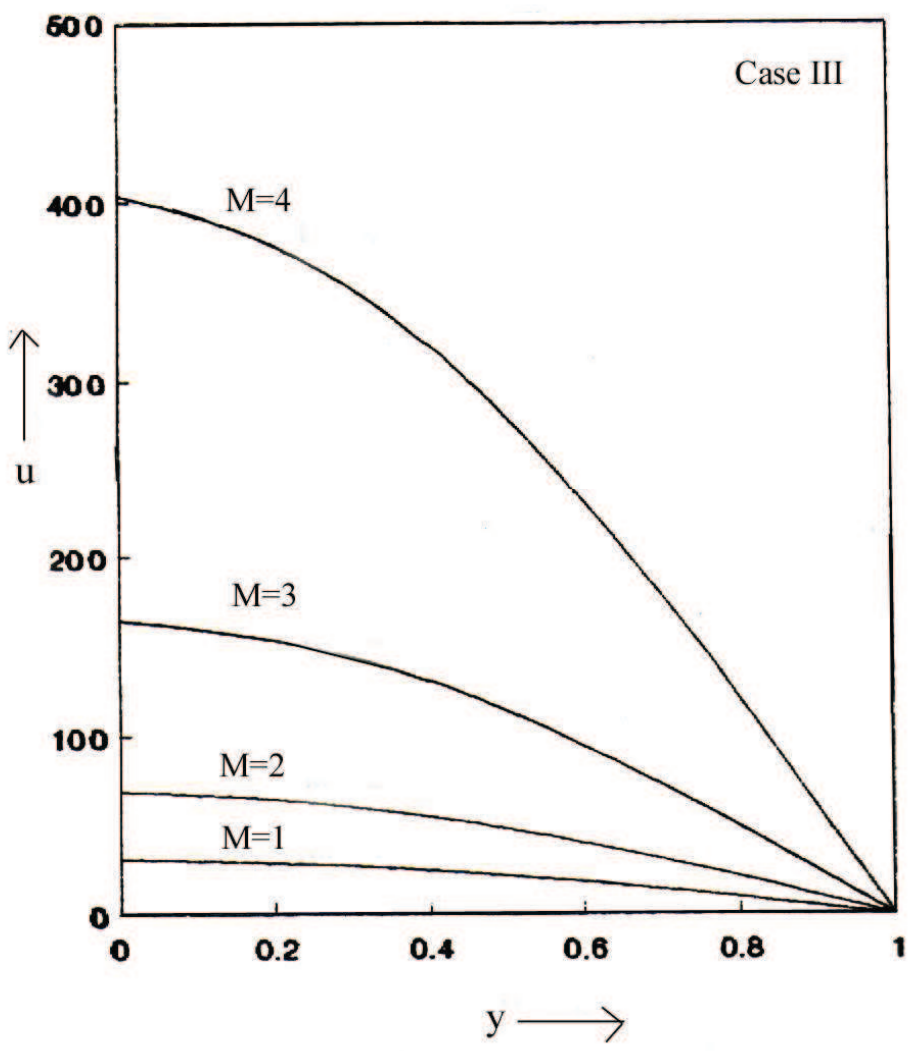

Fig.7 u against y for different M

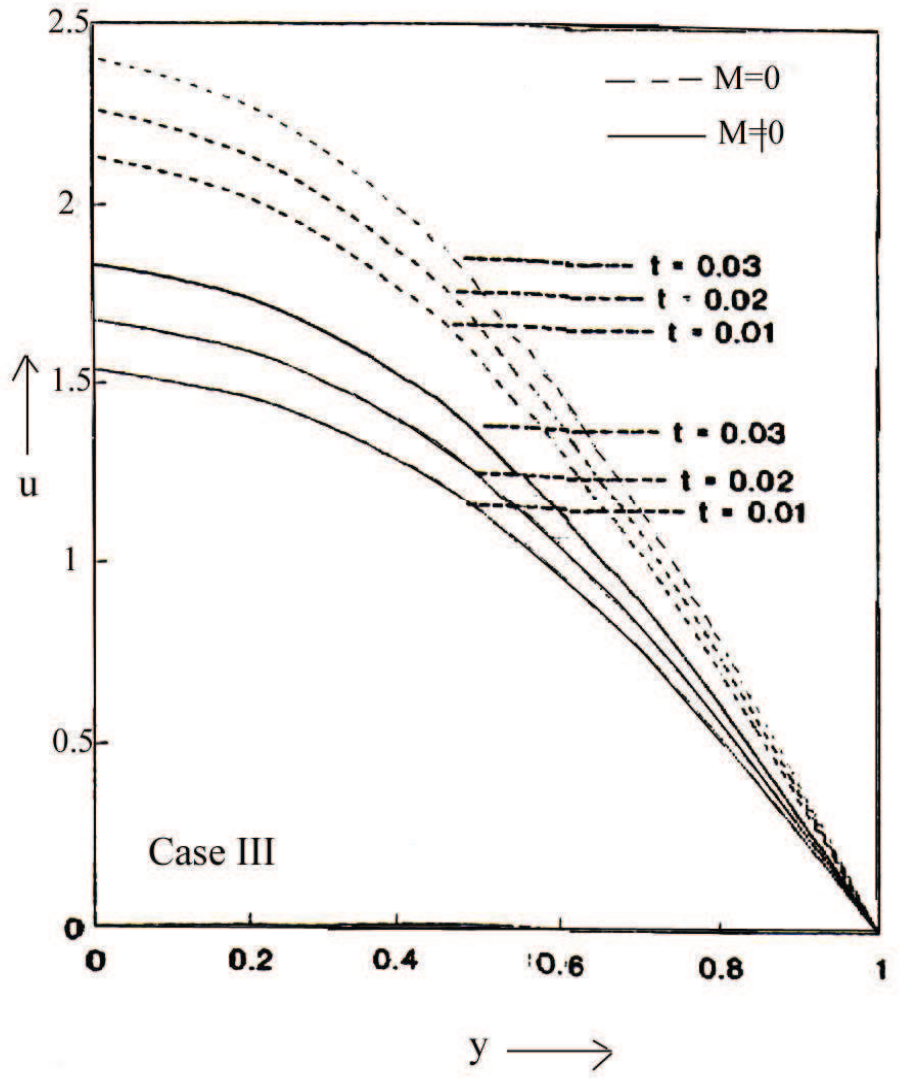

Fig.8 u against y for different t

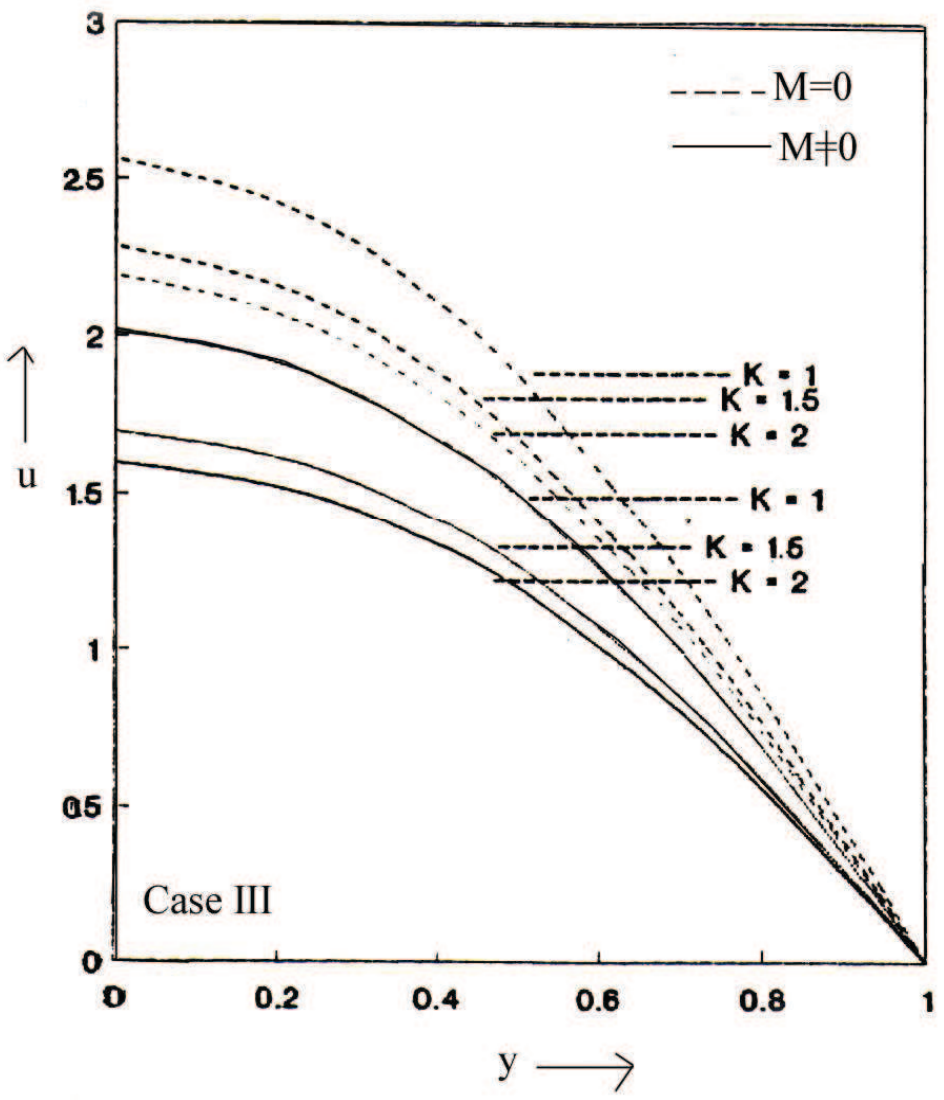

Fig.9 u against y for different K

**Table I**

**Case I U against y for different R**

| R | M | y=0 | 0.2 | 0.4 | 0.6 | 0.8 | 1 |
|---|---|---|---|---|---|---|---|
| 2 | 1 | 30.1757 | 28.72465 | 24.47127 | 17.8251 | 9.406355 | 0 |
| 2 | 0 | 14.51567 | 13.83189 | 11.8035 | 8.622231 | 4.568540 | 0 |
| 4 | 1 | 30.16235 | 28.72465 | 24.47127 | 17.8251 | 9.406355 | 0 |
| 4 | 0 | 14.50283 | 13.83189 | 11.8035 | 80622231 | 4.56854 | 0 |
| 6 | 1 | 30.14952 | 28.72465 | 24.47127 | 17.8251 | 9.406355 | 0 |
| 6 | 0 | 14.48990 | 13.83189 | 11.8035 | 8.622231 | 4.56854 | 0 |
| 8 | 1 | 30.13668 | 28.72465 | 24.47127 | 17.8251 | 9.406355 | 0 |
| 8 | 0 | 14.47716 | 13.83189 | 11.8035 | 8.622231 | 4.56854 | 0 |
| 10 | 1 | 30.12385 | 28.72465 | 24.47127 | 17.8251 | 9.40635 | 0 |
| 10 | 0 | 14.46432 | 13.83189 | 11.8035 | 8.622231 | 4.56854 | 0 |

**Table II**

**Case II U against y for different R**

| R | M | y=0 | 0.2 | 0.4 | 0.6 | 0.8 | 1 |
|---|---|---|---|---|---|---|---|
| 2 | 1 | 34.14086 | 32.50528 | 27.71082 | 20.20846 | 10.68239 | 0 |
| 2 | 0 | 17.33922 | 16.52671 | 14.12038 | 10.33622 | 5.493201 | 0 |
| 4 | 1 | 34.12803 | 32.50528 | 27.71082 | 20.20846 | 10.68239 | 0 |
| 4 | 0 | 17.32638 | 16.52671 | 14.12038 | 10.33622 | 5.493201 | 0 |
| 6 | 1 | 34.11519 | 32.50528 | 27.71082 | 20.20846 | 10.68239 | 0 |
| 6 | 0 | 17.3135 | 16.52671 | 14.12038 | 10.33622 | 5.493201 | 0 |
| 8 | 1 | 34.10235 | 32.50528 | 27.71082 | 20.20846 | 10.68239 | 0 |
| 8 | 0 | 17.30071 | 16.52671 | 14.12038 | 10.33622 | 5.493201 | 0 |
| 10 | 1 | 34.08952 | 32.50528 | 27.71082 | 20.20846 | 10.68239 | 0 |
| 10 | 0 | 17.282788 | 16.52671 | 14.12038 | 10.33622 | 5.493201 | 0 |

**Table III**

**Case III U against y for different R**

| R | M | y=0 | 0.2 | 0.4 | 0.6 | 0.8 | 1 |
|---|---|---|---|---|---|---|---|
| 2 | 1 | 31.2881 | 29.79183 | 25.40008 | 18.52646 | 9.795776 | 0 |
| 2 | 0 | 15.78622 | 15.04873 | 12.86061 | 9.417838 | 5.008056 | 0 |
| 4 | 1 | 31.27597 | 29.79183 | 25.40008 | 18.52646 | 9.795776 | 0 |
| 4 | 0 | 15.77338 | 15.04873 | 12.86061 | 9.417838 | 5.008056 | 0 |
| 6 | 1 | 31.26314 | 29.79183 | 25.40008 | 18.52646 | 9.795776 | 0 |
| 6 | 0 | 15.76055 | 15.04873 | 12.86061 | 9.417838 | 5.008056 | 0 |
| 8 | 1 | 31.2503 | 29.79183 | 25.40008 | 18.52646 | 9.795776 | 0 |
| 8 | 0 | 15.74771 | 15.04873 | 12.86061 | 9.417838 | 5.008056 | 0 |
| 10 | 1 | 31.23747 | 29.79183 | 25.40008 | 18.52646 | 9.795776 | 0 |
| 10 | 0 | 15.73487 | 15.04873 | 12.86061 | 9.417838 | 5.008056 | 0 |